\documentclass[12pt,reqno]{article}
\usepackage{geometry,verbatim}
\usepackage{caption}
\usepackage{subcaption}
\usepackage{eurosym}
\usepackage{dirtytalk}
\usepackage[round]{natbib}
\usepackage{amsmath,amsthm,amssymb,mathtools}
\usepackage{ulem,xpatch}
\xpatchcmd{\sout}
  {\bgroup}
  {\bgroup}
  {}{}
\usepackage{cancel}

\usepackage{dsfont}
\usepackage{url}
\usepackage{lscape}
\usepackage{color}
\usepackage{setspace}
\usepackage{booktabs}
\usepackage[flushleft]{threeparttable}
\usepackage{pdfpages}
\usepackage{booktabs}
\usepackage{threeparttable}
\usepackage{rotating,adjustbox}
\usepackage{multicol}
\usepackage{multirow}
\usepackage{array}
\usepackage{subcaption}
\newcolumntype{L}[1]{>{\raggedright\let\newline\\\arraybackslash\hspace{0pt}}m{#1}}
\newcolumntype{C}[1]{>{\centering\let\newline\\\arraybackslash\hspace{0pt}}m{#1}}
\newcolumntype{R}[1]{>{\raggedleft\let\newline\\\arraybackslash\hspace{0pt}}m{#1}}
\usepackage{amssymb}
\usepackage{pdflscape}
\usepackage{hyperref}
\usepackage{enumitem}
\hypersetup{colorlinks=true, linkcolor=red,          
    citecolor=blue,        
    filecolor=magenta,      
    urlcolor=cyan           
}
\usepackage{tikz}
\usepackage{tikz-qtree}
\usetikzlibrary{arrows,shapes,positioning,shadows,trees,calc, automata}
\tikzset{every node/.style={scale=0.6, > = stealth, node distance = 3cm, 
    on grid, auto, font=\small}, 
every state/.style={draw = black!10,fill = black!10,minimum size=4mm},
tt/.style ={circle,draw = red,fill = red!40,minimum size=4mm},       
removed/.style = {circle, draw = white,fill = white, minimum size=4mm},
t1/.style ={circle,draw = blue,fill = blue!40,minimum size=4mm},
unattractive/.style ={dashed,circle, draw = black,fill = black!10,minimum size=4mm}
}

\theoremstyle{plain}

\newtheorem{lemma}{Lemma}
\newtheorem{proposition}{Proposition}
\theoremstyle{definition}
\newtheorem{definition}{Definition}

\theoremstyle{remark}
\newtheorem{example}{\textbf{Example}}

\DeclareGraphicsExtensions{.eps,.ps,.eps.gz,.ps.gz,.eps.Z}

\usepackage{datetime}
\newdateformat{monthyeardate}{%
\monthname[\THEMONTH], \THEYEAR}
\usepackage{dashbox}

\begin{document}
\title{Respecting priorities versus respecting preferences in school choice:
When is there a trade-off?\thanks{Funding
from FNRS PDR grant 23671793, ANII FMV-1-2021-1-166576, and Tore Browaldhs Stiftelse BFh18-0007 is gratefully acknowledged. We thank Péter Biró, Lars Ehlers, Bingling Gong, Bettina Klaus, Alexey Kushnir, Peng Liu, David Manlove, Antonio Miralles, Antonio Nicolò, Parag Pathak, Olivier Tercieux, Camille Terrier, Chris Wallace and seminar and conference audiences for comments and suggestions. The simulations were enabled by resources in project SNIC 2022/22-675 and NAISS 2023-22-700 provided by the Swedish National Infrastructure for Computing (SNIC) at UPPMAX, partially funded by the Swedish Research Council through grant agreement no. 2018-05973.
}}
\author{Estelle Cantillon\thanks{FNRS, Universit\'{e} Libre de Bruxelles (ECARES) and CEPR, email: Estelle.Cantillon@ulb.be.},
Li Chen\thanks{Corresponding author, Shanghai International College of Intellectual Property, Tongji University and Department of Economics, University of Gothenburg, email: lichen.cn@outlook.com.}  
\hspace{0.15cm}and Juan Pereyra\thanks{Department of Economics, Universidad de Montevideo, Uruguay, email: jspereyra@um.edu.uy.}}
\date{September 2024}
\maketitle

\begin{abstract}
\noindent A classic trade-off that school districts face when deciding which matching algorithm to use is that it is not possible to always respect both priorities and preferences. The student-proposing deferred acceptance algorithm (DA) respects priorities but can lead to inefficient allocations. We identify a new condition on school choice markets under which DA is efficient. Our condition generalizes earlier conditions by placing restrictions on how preferences and priorities relate to one another only on the parts that are relevant for the assignment. Whenever there is a unique allocation that respects priorities, our condition captures all the environments for which DA is efficient. We show through stylized examples and simulations that our condition significantly expands the range of known environments for which DA is efficient. We also discuss how our condition sheds light on existing empirical findings.
\medskip{}

\noindent \textit{Keywords:} Matching, envyfreeness, fairness, efficiency, priorities, preferences, mutually best pairs. 
\medskip{}

\noindent \textit{JEL Codes:  C78, D47, D78, D82}
\end{abstract}
\textit{\newpage}

\section{Introduction}
Many countries promote parental choice for schools. How choice is actually implemented greatly varies across countries and school districts and is often hotly debated by stakeholders. A cornerstone of this debate is the choice of the algorithm used to allocate students to schools, when capacity is limited. Indeed, different algorithms will typically result in different final allocations, even when submitted preferences and priorities are the same. 

In their classic paper, \cite{abdulkadirouglu2003school} identify desirable properties of matching algorithms in the school choice context. These include efficiency, which can be interpreted as a measure of the extent to which the algorithm respects students' preferences, and envyfreeness, a measure of the extent to which priorities are respected. Unfortunately, no algorithm exists that always produces an envyfree and efficient final allocation \citep{balinski1999tale}, and there is a trade-off between respecting students' preferences and respecting schools' priorities. Two strategyproof mechanisms satisfy one
of the desiderata: the student-proposing deferred acceptance (DA) algorithm,\footnote{When not specified, we refer to the student-proposing DA algorithm as DA in short.} originally proposed by \citet{gale1962college}, always yields an envyfree allocation; the top trading cycles (TTC) algorithm, first described by \cite{shapley1974cores}, is efficient. Other non-strategyproof mechanisms are used in practice. They are often variants of the school-proposing DA, which always produces an envyfree matching but is not efficient \citep{roth1984misrepresentation}, or of the immediate acceptance (IA) algorithm, also known as the Boston mechanism, which produces an envyfree allocation in equilibrium \citep{ergin2006games}, but may not be efficient.  

\begin{table}[htp]

\caption{Tensions between efficiency and envyfreeness across school districts}\label{tab:Evidence}
\footnotesize 
\centering 
\scalebox{0.9}{
\begin{tabular}{L{4cm}L{3cm}L{2cm}L{2cm}L{4cm}}
\toprule
School district & Algorithms compared & $\%$ students with Pareto improving trade & $\%$ students with justified envy & Special features \\  \midrule

Boston, all levels \citep{atila2006changing,pathak2017really} & Student-proposing DA, TTC & & 6.8 &   Sibling and neighborhood priority\\ \midrule
Budapest, secondary \citep{biro2012matching, ortega2022improving} &  Student-proposing DA, TTC & &  64 &   Combination of school grades, centralized exam and own school test/interview\\ \midrule
Ghent elementary (own source)&  School-proposing DA, TTC &  $<1$ & 9.2 &  Sibling and staff priority, distance as tie-breaker\\ \midrule
New Orleans - elementary to middle school
\citep{abdulkadirouglu2020efficiency}
&  School-proposing DA, TTC &   & 13  &  Sibling priority, catchment area\\ \midrule
New York, high school \citep{abdulkadirouglu2009strategy} & Student-proposing DA, TTC &  $\geq 5.45$ & 44  & Mix of schools and of priority and ranking criteria \\ 

\bottomrule
\end{tabular}
}
\medskip  

\begin{minipage}{0.95\linewidth}
  \textbf{Notes:} There are no commonly accepted measures of the efficiency and envyfreeness trade-off. We consider two ordinal measures. A first measure is the fraction of students in the student-proposing DA with an available Pareto improving trade. A student is said to have a Pareto improving trade if there exists one or more students, and exchanges of seats among them, such that all are better off based on submitted preferences. A second measure, taking the TTC allocation as a starting point, is the fraction of students who have justified envy, i.e. they are not accepted at a school that they prefer to their assigned school, even though a student with lower priority got in (\cite{dougan2022robust} provide foundations for this measure). 
\end{minipage}
\end{table}

How big the trade-off is in practice is an empirical question, and existing evidence is mixed. Table \ref{tab:Evidence} summarizes some of the available evidence. \cite{abdulkadirouglu2009strategy} and \cite{che2019efficiency} have documented a significant trade-off between respecting students' preferences and schools' priorities in the NYC High School markets. In Budapest, \cite{ortega2022improving} have documented large violations of priorities when using TTC. Using estimated preferences, \cite{calsamiglia2020structural} and \cite{de2023performance} also find significant differences across algorithms. This contrasts with the evidence from the Boston Public Schools system where \cite{atila2006changing} and \cite{pathak2017really} have found very little difference between the outcomes of DA and TTC. Likewise, we were able to look at data from the allocation of seats in elementary schools in the city of Ghent (Belgium) and found the school-proposing DA close to be efficient. Evidence from New Orleans is somewhere in between.

These examples raise the question of when the choice of algorithms actually matters: When should a school district spend time weighting the choice among algorithms? When is the choice of the algorithm a second-order issue?

We identify a new condition on preferences and priorities under which there is no trade-off between the two goals: there is an envyfree and efficient allocation, and in fact it is also unique. Our condition, Generalized Mutually Best Pairs (GMBP), generalizes existing conditions identified in the literature. Like some of the existing conditions, it seeks to capture the degree to which priorities and preferences are congruent, but it restricts attention to the parts of the preferences and priorities that actually matter for the allocation (we call it the \textit{simplified market}). Some students who have priority in a school may never need to consider it because they can access a preferred school for sure. Reversely, some schools that students have listed in their preferences may not be attainable anyways. 
Roughly speaking, our GMBP condition is satisfied if, in the simplified market, we can \textit{sequentially} match students to the best school in their preference lists for which their priority qualifies them for one of the available seats. If a market satisfies the GMBP condition, then there is no trade-off between efficiency and envyfreeness. There is a unique envyfree allocation and it is efficient. It is reached
at equilibrium by the student-proposing DA, the school-proposing DA, the IA algorithm or any other mechanism that produces an envyfree allocation. Moreover, if we can sequentially match students to their best available schools in the original market (without simplifying), then all the algorithms -- including TTC -- produce the same allocation that is both efficient and envyfree. 

Whenever there is a unique envyfree allocation, our GMBP condition is not only sufficient but also necessary for this allocation to also be efficient. This means that other environments where efficiency and envyfreeness are compatible are also environments where there are multiple envyfree allocations and where, therefore, the choice of the algorithm matters.

We discuss how our GMBP condition naturally arises in existing school choice environments. An expanding empirical literature has shown that parents typically value school quality and proximity \citep{burgess2015parents,abdulkadirouglu2017welfare,fack2019beyond}. We show that such preferences, when combined with priorities based on distance, meet our GMBP condition. Likewise, to the extent that preferences for academic quality correlate with academic performance, priorities based on academic performance will also meet the condition. We show through simulations that our condition significantly expands the set of known environments for which there is no trade-off between efficiency and envyfreeness. Our results shed light on the differences in performance across school environments described in Table \ref{tab:Evidence}.

\textbf{Positioning within the literature.} A number of papers have explored the trade-off between envyfreeness and efficiency in school choice markets, starting from the seminal papers by \cite{balinski1999tale} and \cite{abdulkadirouglu2003school}. The DA and TTC algorithms are natural starting points here: DA maximizes efficiency among algorithms that produce envyfree outcomes \citep{gale1962college, balinski1999tale}, and TTC performs well (and under some circumstances best) on envyfreeness, within the class of efficient and strategyproof mechanisms \citep{abdulkadirouglu2020efficiency, dougan2022robust, dur2022characterization}.  When TTC does not involve any violation of priorities, it yields the same outcome as DA.

One approach has been to identify domain restrictions on the set of priorities such that DA is efficient \citep{ergin2002efficient, ehlers2010efficient, erdil2019efficiency, ishida2019two} or TTC yields the same outcome as DA \citep{kesten2006two, ishida2019two}, for any preference profile.\footnote{The second question is more restrictive as DA can be efficient without producing the same allocation as TTC, whereas TTC yields the same allocation as DA when it is envyfree.} \cite{heo2019preference} explores the flip-side question of domain restrictions on preferences such that DA is efficient or TTC yields the same outcome as DA, for any priority profile.

Of course, whether efficiency and envyfreeness conflict depends on the \textit{combination} of preferences and priorities. Moreover, in practice, priorities are not entirely independent of preferences and are instead a partial reflection of what school districts (or schools) view as legitimate preferences. For example, many school districts give priorities to siblings or to students living close to schools (cf. Table \ref{tab:Evidence}), and parents often prefer to send their children to nearby schools or to the school where a sibling is already educated, everything else equal (see e.g. \citealp{burgess2015parents, abdulkadirouglu2017welfare, harris2023schools}).

Our GMBP condition captures this insight and generalizes existing conditions, placed on how preferences and priorities relate, known to guarantee that DA is efficient, such as \cite{salonen2018mutually}'s single peakedness, \cite{clark2006uniqueness}'s no crossing condition or \cite{reny2021simple}'s student-oriented preferences.\footnote{\cite{clark2006uniqueness} originally solely focused on the uniqueness of stable matching in one-to-one two-sided matching but \cite{salonen2018mutually} show that his condition also ensures that DA is efficient (and that TTC yields the same outcome as DA).} 

The trade-off between efficiency and envyfreeness is conceptually connected to the issue of uniqueness of stable matchings which has been explored in one-to-one two-sided markets \citep{alcalde1994exchange, eeckhout2000uniqueness,clark2006uniqueness, niederle2009decentralized, legros2010co, romero2013acyclicity,lee2014efficiency, gutin2023unique}. Intuitively, given the lattice structure of stable matchings \citep{knuth1997stable}, uniqueness of stable matchings suggests that preferences on both sides (priorities and preferences in the school choice context) are “sufficiently compatible” that starting from one side or the other when using deferred acceptance does not impact the final outcome. Methodologically, our procedure of removing irrelevant choices of students to define the simplified market, is related to the matching problem reduction of \cite{gutin2023unique}. The main difference is that our simplification process only removes irrelevant choices from students' preferences and update schools' priorities accordingly. We do not remove irrelevant students for schools, because doing so impedes efficiency. 

Though uniqueness of stable matchings does not in itself guarantee efficiency (in a school choice context) nor the equivalence between DA and TTC, we argue in Section \ref{relevance} that most existing conditions that have been identified to imply uniqueness rely, in their definition or their proof, on some form of mutually best pairs condition. They therefore, once appropriately extended to one-to-many matching contexts, also imply that DA is efficient. Our proposition \ref{multiple} makes this connection between the two strands of the literature precise: our GMBP condition characterizes all the environments for which there is a unique stable matching \textit{and} this stable matching is efficient. 

The rest of the paper is organised as follows. Section \ref{sec:model} describes our model. Section \ref{sec:gmbp} introduces our GMBP condition, shows how it generalizes earlier conditions based on the same idea, and establishes our main theoretical results. Section \ref{relevance} then revisits stylized environments used in the literature to explore to what extent our condition captures a larger set of those. Section \ref{sec:simulation} does the same using numerical simulations instead. So, the question is no longer whether ``all environments of a specific kind" meet the condition but ``what fraction of environments of a specific kind" meets the condition. Section \ref{sec:conclusion} concludes. 

\section{Model}\label{sec:model}

A wide variety of algorithms and procedures are used in practice to match students to schools. We focus on \textit{direct priority-based} mechanisms which elicit preference rankings from students (this is the ``direct" part) and use priorities to assign students to schools when demand exceeds capacity (the ``priority" part).  

Let $I$ denote the set of students ($|I|=n$), and $S$ the set of schools ($|S|=m$). Let school $s_{m+1}$ represent the outside option for the students. A market is defined by students' preferences over schools, schools' priorities over students and schools'
capacities, and is denoted by $\mathcal{E}=(\succ,\ P,\ q)$, where $\succ$
are students' preferences over $S\cup\{s_{m+1}\}$, $P$ are schools'
priorities, and $q=(q_{1},...,q_{m})\in\mathbb{\mathbb{\mathbb{N}}}^{m}$
are school capacities (without loss of generality, we let the capacity
of $s_{m+1}$ to be equal to $n)$. We assume that preferences and
priorities are strict linear orders (no indifference). In addition, the priority order of each school only contains those students who consider the school as acceptable, i.e. preferred to their outside options.\footnote{This assumption is without loss of generality as we will focus on individually rational allocation.}  

An allocation is a mapping $\mu:\:I\rightarrow S\cup\{s_{m+1}\}$ that describes the school to which students are assigned, with the understanding that $\mu(i)=s_{m+1}$ means that student $i$ is not assigned. An allocation
is feasible if it does not allocate more students to a school than
its capacity, $|\mu^{-1}(s)|\leq q_{s}$ , for every $s \in S$. A feasible allocation $\mu$ is \textit{(Pareto) efficient} if it is not Pareto dominated by any other feasible allocation $\mu'$, that is, if there is no $\mu'$ such that $\mu'(i) \succeq_i 
\mu(i)$ for all $i\in I$ and $\mu'(i) \succ_i
\mu(i)$ for some $i\in I$. A feasible allocation is blocked by a pair $(i,s)$ if $i$ prefers $s$ to her assignment $\mu(i)$, and either $s$ has some empty seats under $\mu$, or there is a lower priority student $j$ who is assigned to $s$ under $\mu$, that is, formally, 
$s\succ_i \mu(i)$, and $|\mu^{-1} (s)| <q_s$ or $iP_sj$ for some $j \in \mu^{-1} (s)$. An allocation is said to be \textit{envyfree} or \textit{stable} if it is not blocked by a pair.\footnote{Given our modeling choice for the outside option, this definition implies that an envyfree allocation is also individually rational since students can always block with their outside option.} 

A direct priority-based mechanism is a function that maps student preferences $\succ$, school capacities $q$ and priorities $P$ into a feasible allocation. Denote $\phi_i(\succ_i,\succ_{-i})$ the school allocation of student $i$ when the submitted preferences are ($\succ_i,\succ_{-i}$). A mechanism $\phi$ is \textit{strategyproof} if it is a weakly dominant strategy for all students to submit their true preferences, i.e. if for all $i$, $\phi_i(\succ_i,\succ_{-i})\succeq_i \phi_i(\succ'_i,\succ_{-i})$, for every $\succ_i,  \succ'_i$, and every $\succ_{-i}$.

We first consider two mechanisms at the center of the debate between envyfreeness and efficiency. The first one is the student-proposing DA, first proposed by \cite{gale1962college}. The mechanism is strategyproof and always produces an envyfree allocation based on the submitted preferences and priorities. The allocation it produces Pareto dominates all other envyfree allocations \citep{gale1962college, balinski1999tale}.   
The second algorithm, TTC, was first described in \cite{shapley1974cores} and adapted to the school choice problem by \cite{abdulkadirouglu2003school}. The mechanism is strategyproof and always produces an efficient allocation based on submitted preferences. However, it may violate priorities and thus does not guarantee an envyfree allocation.

The student-proposing DA and TTC serve as natural benchmarks to measure the tension between envyfreeness and efficiency: the student-proposing DA maximizes efficiency among algorithms that produce envyfree allocations and TTC performs well (and under some circumstances best) on envyfreeness. Specifically, \cite{abdulkadirouglu2020efficiency} and \cite{dougan2022robust} have shown that TTC minimizes envy within the class of efficient and strategyproof mechanisms when all schools have unit capacity. When schools have capacities larger than one, \cite{dur2022characterization} show that TTC is the only efficient and strategyproof mechanism among the class of mechanisms that satisfy a slightly weaker version of envyfreeness.\footnote{\cite{morrill2015two} and \cite{hakimov2018equitable} have proposed variants of \cite{abdulkadirouglu2003school}'s TTC version for multiple units that reduce envy. Because none of these variants would change our results, we work with the better known \cite{abdulkadirouglu2003school}'s variant.}  

Additionally, two common mechanisms used in practice are the school-proposing DA and the Immediate Acceptance (IA) algorithm. Unlike its student-proposing counterpart, the school-proposing DA offers seats to students in order of school priorities. It always produces an envyfree allocation but is neither efficient nor strategyproof: in the Nash equilibrium of the school-proposing DA, students may be tempted to misreport their preferences to get a better (still envyfree) allocation \citep{roth1982economics}.

Like student-proposing DA, the IA mechanism starts with students' preferences but it allocates seats first to first choices (possibly using school priorities if demand is higher than the number of seats), before  considering second choices. The IA mechanism is efficient based on submitted preferences but it is not strategyproof: some students may have incentives to submit preferences different from their true preferences in order to get a preferred assignment. \cite{ergin2006games} show that the set of Nash equilibrium outcomes is equal to the set of feasible envyfree allocations. 

In the rest of the paper, we assume that students play the weakly dominant strategy of truth telling under both student-proposing DA and TTC and that they play Nash equilibrium strategies under the school-proposing DA and the IA mechanisms. All mechanisms are described formally in the Appendix. 

\section{The Generalized Mutually Best Pairs condition}\label{sec:gmbp}

We are interested in analyzing the set of markets -- conditions on priorities and preferences -- for which there is no conflict between envyfreeness and efficiency. Since the DA allocation Pareto dominates all other envyfree allocations, the most direct formal translation of this question is to ask when DA is efficient. A slightly more demanding requirement is to ask when DA yields the same allocation as TTC. This requirement is more demanding because DA can be efficient, and yet yield a different allocation than TTC. In that case, TTC is efficient (as always) but not envyfree. 

Our condition builds on \cite{eeckhout2000uniqueness}'s sufficient condition for uniqueness in one-to-one two-sided matching environments. Specifically, in a context where all schools have unit capacity, \cite{eeckhout2000uniqueness}'s condition comes down to requiring that there exists a reordering of schools and students such that (unit capacity) schools and students are \textbf{mutually best pairs}, i.e.  for each $i \in I$, $s_i \succ_{i} s_k $ for all $k>i$ and $i P_{s_{i}} k$ for all $k>i$. 

In an environment with unit capacity, this condition trivially ensures that DA is efficient and leads to the same allocation as TTC. To see this, consider first student 1 and school 1. School 1 is student 1's top choice and since they have the highest priority in that school, they will be assigned to that school under DA.\footnote{We use ``they'' as the gender-neutral third person singular pronoun.} Next, consider student 2. They may prefer school 1 to school 2 but under the mutually best pairs condition, school 2 is their top or second top (after school 1) preferred school. Hence, since student 2 is also school 2's preferred student (except possibly for student 1 who is already assigned), student 2 will be assigned to school 2 under DA. As the process continues, it is clear that, under DA, student $i$ will be allocated to school $s_i$. This outcome is also the outcome that arises from TTC by iterated elimination of unit length cycles (school 1 pointing to student 1 and the reverse, etc). 

\cite{eeckhout2000uniqueness} finds that this condition is sufficient for the set of envyfree allocations to be singleton and we have just seen that it is also sufficient for DA to be efficient and for DA to yield the same allocation as TTC. 
The next example suggests that this condition is over-restrictive, even in the unit capacity school context.
\medskip 

\begin{example}\label{example:truncation}
  Consider a market with unit-capacity schools. We describe students' preferences by listing, for each student, their acceptable schools in their preferences orders. Similarly for schools' priorities. The preferences and priorities are as follows:
  
  \[
  \begin{array}{cccc}
  i_1: & s_1 & s_3 & s_2 \\
  i_2: & s_2 & s_1 & \\
  i_3: & s_3 &     &
  \end{array}
  \hspace{50pt}
  \begin{array}{cccc}
  s_1: & i_2 & i_1 &  \\
  s_2: & i_1 & i_2 &  \\
  s_3: & i_1 & i_3 &
  \end{array}
  \]
  \smallskip 

It is easily checked that DA and TTC yield the same allocation, $(i_1, s_1), (i_2, s_2), (i_3, s_3)$, where $(i,s)$ means that student $i$ is assigned to school $s$, and that there is a unique envyfree allocation. Yet,  this market does not satisfy \cite{eeckhout2000uniqueness}'s condition because there is no mutually best pair to start with. 

Now consider school $s_2$ in student $i_1$'s preferences. School $s_2$ is an irrelevant choice for student $i_1$ since that student is ranked first by school $s_3$ and they prefer school $s_3$ to school $s_2$. Hence, student $i_1$ will never be assigned to school $s_2$ in any envyfree allocation and we might as well remove them from the priority of school $s_2$.  

Building on this intuition, we can define for every school market an associated ``simplified school market" where students' preferences are truncated at the best school they can get for sure (their ``safe school") and priorities are updated accordingly. In the example above this gives, after two additional rounds of truncations: 

\[
  \begin{array}{cccc}
  i_1: & s_1  \\
  i_2: & s_2  \\
  i_3: & s_3 &     & \\
  \end{array}
  \hspace{50pt}
  \begin{array}{cccc}
  s_1: & i_1  \\
  s_2: & i_2  \\
  s_3: & i_3 &
  \end{array}
  \]
  \smallskip 

This simplified market satisfies \cite{eeckhout2000uniqueness}'s mutually best pairs condition. We will formally show in Lemma \ref{lemma:same} below that, by construction, it admits the same set of envyfree allocations as the original market. Therefore, we can leverage the mutually best pairs condition to conclude that the original market admits a unique envyfree allocation and that DA is efficient.\footnote{In general, truncation will change the outcome of TTC so the fact that TTC yields the same outcome as DA in the simplified market does not necessarily mean that it will do so in the original market.} $\square$
\end{example}

\medskip 
Following Example \ref{example:truncation}, we will consider, from now on, simplified markets where irrelevant choices have been removed from students' rank order lists through iterated truncations at the best ``safe school", and priority lists have been updated accordingly.

To formalize the idea of simplified markets, we first define the concept of irrelevant school for a student. 
\begin{definition}[\textbf{Irrelevant school}]
    Given a school choice market $\mathcal{E}=(\succ,\ P,\ q)$, we say that school $s$ is irrelevant for student $i$ if there exists a school $s'\neq s$ such that $s'\succ_i s$, and $\vert j \in I : j P_{s'} i \vert < q_{s'}$.
\end{definition}

In words, school $s$ is irrelevant for a student if there is another school which this student prefers and which they are sure to get because they are high enough in that schools' priority order. Note that a student will never be assigned to an irrelevant school in an envyfree allocation.

We then define the following process of \textit{iterative elimination of irrelevant schools}. 
\medskip 

Step 1: For each student $i$, find all their irrelevant schools. If no student has an irrelevant school, stop the process. Otherwise, for each student delete the irrelevant schools from their preferences, and delete the student from the priority list of each irrelevant school. 

Step $k\geq 2$: In the new market with the modified preferences and priorities, repeat Step 1.

The process finishes when no student has an irrelevant school. 
\medskip 

\begin{definition}[\textbf{Simplified market associated with $\mathcal{E}$}]
\label{def:simple} Consider any school choice market $\mathcal{E}=(\succ,\ P, \ q)$. Its associated simplified market is given by the outcome of the iterative elimination of irrelevant schools and is denoted $\mathcal{E^*}=(\succ^*,P^*, q)$, where $\succ^*$ corresponds to the truncation of $\succ$ before the highest irrelevant school and $P^*$ corresponds to a selection of $P$ to the students that have the schools on their truncated preference lists.
\end{definition}

Because the simplification process only removes irrelevant schools from students' preferences (and the students from the priorities of their irrelevant choices), the set of envyfree allocations in the simplified market is the same as in the original market as the next lemma shows.

\begin{lemma}\label{lemma:same}
The sets of envyfree allocations in $\mathcal{E}$ and its associated simplified version, $\mathcal{E^*}$, are the same.  
\end{lemma}

\begin{proof} We first show that all envyfree allocations in $\mathcal{E}$ are also envyfree in $\mathcal{E^*}$. Consider the envyfree allocation $\mu$ in $\mathcal{E}$ and suppose it is not envyfree in $\mathcal{E^*}$. This means there exists $(i,s)$ such that $s\succ_i^* \mu(i)$, and $|\mu^{-1} (s)| <q_s$ or $i P_s^* j$ for some $j \in \mu^{-1} (s)$. But this means that $s\succ_i \mu(i)$ since $\succ^*$ is a truncation of $\succ$ and either $|\mu^{-1} (s)| <q_s$ or $i P_s j$ for some $j$ (since $P^*$ is a selection of $P$). A contradiction. 

Consider now $\mu$, an envyfree allocation in $\mathcal{E^*}$, and suppose it is not envyfree in $\mathcal{E}$. This means there exists $(i,s)$ such that $s\succ_i \mu(i)$, and $|\mu^{-1} (s)| <q_s$ or $i P_s j$ for some $j \in \mu^{-1} (s)$. We claim that this implies that $s\succ_i^* \mu(i)$, and $|\mu^{-1} (s)| <q_s$ or $i P_s^* j$ for some $j \in \mu^{-1} (s)$. 

To see this, first consider $\succ^*$, and note that the only schools that were removed from $\succ$ during the simplification process are schools that were irrelevant (i.e. not envyfree). Therefore, if $\mu(i)$ is part of an envyfree allocation, then it was not removed from student $i$'s rank order list and neither were any of the schools preferred to $\mu(i)$ according to $\succ_i$ (simplification only results in a truncation). Hence, $s\succ_i^* \mu(i)$. 

Next, note that $i P_s j$ implies that $i P_s^* j$ because only those students who view a school as irrelevant are removed from that school's priority list during the simplification process. Therefore, we conclude that $\mu$ cannot be an envyfree allocation in $\mathcal{E^*}$, a contradiction.
\end{proof}

Our second example illustrates the added difficulty arising from allowing schools to have multiple seats. 

\begin{example}\label{example:multiple-capacity}
  Consider the following market with preferences, capacities and priorities as follows:
   \[
  \begin{array}{cccc}
  i_1: & s_2 & s_1& \\
  i_2: & s_1 & &  \\
  i_3: & s_1 & s_2 &\\
  i_4: & s_1 & s_2 & s_3\\
  \end{array}
  \hspace{50pt}
  \begin{array}{cccccc}
  s_1: & i_1 & i_2 & i_3 & i_4  &   \text{(capacity = 2)} \\
  s_2: & i_2 & i_1 & i_4 & i_3    & \text{(capacity = 1)} \\
s_3: &  i_4 & & & & \text{(capacity = 1)} 

  \end{array}
  \]
  \smallskip  
  
Note first that the school choice environment cannot be further simplified: there are no irrelevant choices in students' rank order lists. Furthermore, this market does not satisfy the mutually best pairs condition, even if we divide the schools into mini-schools of unit capacity that inherit the priorities of the original schools as is typically done (see e.g. \citealp{abdulkadirouglu2003school}). Yet, DA is efficient and produces the same allocation as TTC, namely, $(i_1, s_2),  (i_2, s_1),  (i_3, s_1),  (i_4,s_3)$.

The issue here is that only school $s_1$'s top priority student is considered when checking the mutually best pairs condition, even though school $s_1$ can admit 2 students. 

One way to address this issue is to consider schools' top $q$ students when identifying mutually best pairs. So, in this case, student $i_2$ and school $s_1$ are mutually best pairs since student $i_2$ is one of school $s_1$'s top 2 students.
Once student $i_2$ is removed from schools' priority lists, student $i_1$ and school $s_2$ become mutually best pairs, leaving student $i_3$ to be mutually best pair with school $s_1$, and finally student $i_4$ and  school $s_3$ are a mutually best pair. $\square$

\end{example}
\medskip

We are now ready to introduce our generalized mutually best pairs condition. We do this in two steps. We first define the sequential mutually best pairs (SMBP) condition, which is the generalisation of Eeckhout's condition to environments with multi-unit capacity.

\begin{definition}[\textbf{Sequential mutually best pairs condition}]\label{def:SMBP}
A market $\mathcal{E}=(\succ,P,\ q)$ satisfies the sequential mutually best pairs condition if there is a reordering of students $(i_1, i_2, ...)$ and an associated list of schools $S$, $(s_{(1)}, s_{(2)}, ...)$, where $s_{(i)} \in S$ stands for the school associated with student $i$ and the same school does not appear more times than its capacity, such that:
\begin{enumerate}
  \item $s_{(1)}\succeq_{i_1} s$ for all $ s \in S \setminus \{s_{(1)}\}$ and $i_1$ is among the top $q_{s_{(1)}}$ students in school $s_{(1)}$'s priority list,
  \item (for $k>1$), $s_{(k)}\succeq_{i_k} s$ for all $s \in S^k = \{s \in S: q_s^k >0\} $, where $q_{s}^k = q_s - \sum_{l=1}^{k-1} 1_{\{s_{(l)} = s\}}$ is the remaining capacity of school $s$ by the time we reach student $i_k$, and student $i_k$ is among the top $q_{s_{(k)}}^k$ students in school $s_{(k)}$'s priorities, among students $i_k, i_{k+1}, ....$.  
\end{enumerate}
\end{definition}

\medskip

In words, a market satisfies the SMBP condition if we can \textit{sequentially} match students to the best school in their preference lists for which their priority qualifies them for one of the remaining seats. The SMBP condition describes environments where \cite{salonen2018mutually}'s Iterated Best Match algorithm converges (produces a non-wasteful matching, in their terminology).\footnote{The Iterated Best Match algorithm runs as follows (\citealp{salonen2018mutually}, p. 44) : 1. Students (schools) report their preferences over schools (students) that are in the market. If there are no mutually best pairs, then stop. If there are mutually best pairs, the mechanism matches permanently all such pairs. Update the preferences and capacities of the agents. Permanently matched students and those schools whose capacity is full leave the market. Also those students (schools) who no
longer have acceptable schools (students) leave the market (this step is repeated if necessary). Go to step 2. 2. If the sets of remaining students and schools with remaining capacities are not empty, repeat step 1. Else stop.}
The condition can be easily checked by looping over students and checking whether students belong to the top priority students of their preferred school (with school capacities adjusted as students get matched to them). 

To illustrate Definition \ref{def:SMBP} with Example \ref{example:multiple-capacity}, we can reorder students and schools as follows: 
\[
\begin{array}{cccc}
  i_2 & i_1 & i_3 & i_4\\
  s_1 & s_2 & s_1 & s_3 \\
  \end{array}
  \]

Our generalized mutually best pairs (GMBP) condition imposes the SMBP condition only on the simplified version of the original market and is thus a relaxation of the SMBP condition.

\begin{definition}[\textbf{Generalized mutually best pairs condition}]\label{def:GMBP}
A market $\mathcal{E}=(\succ,P,\ q)$ satisfies the generalized mutually best pairs condition if its simplified market $\mathcal{E^*}$ satisfies the sequential mutually best pairs condition.
\end{definition}

While the sequential process that underlies the SMBP condition serves to eliminate unrealistic choices in the top of students’ preferences (by removing schools that no longer have capacity by the time a student is considered), the market simplification process used for GMBP removes uninterested students from the top of schools’ priorities. The two actions together help focus on the relevant set of possible allocations. We defer the detailed discussion of which school market environments are likely to satisfy the GMBP condition until Section \ref{relevance}. Let us simply note for now that the condition captures a sense of compatibility between students' preferences and schools' priorities (mutually best pairs), once we account for the feasible choice set of students (the mutually best condition is checked sequentially, on the relevant set of alternatives). We have the following result: 

\begin{proposition}[\textbf{No trade-off between efficiency and envyfreeness}]\label{GMBP}
Suppose $\mathcal{E}=(\succ,\ P,\ q)$ satisfies the generalized mutually best pairs condition. Then, there is a unique envyfree allocation and it is efficient. It is produced by the  student-proposing DA, school-proposing DA and IA. 
\end{proposition}

\begin{proof} 
We first establish that there is a unique envyfree allocation in the simplified market $\mathcal{E^*}$. By Lemma \ref{lemma:same}, this implies that there is a unique envyfree allocation in $\mathcal{E}$ as well.

Consider the allocation produced by the reordering of students and schools associated with the application of the SMBP condition on the simplified market: $(i_1, s_{(1)})$, $(i_2, s_{(2)})$, ..., $(i_n, s_{(n)})$. 

We claim that this is the unique envyfree allocation. Consider student $i_1$. By definition, school $s_{(1)}$ is student $i_1$'s most preferred school and student $i_1$ belongs to the highest $q_{s_{(1)}}$ priority students of school $s_{(1)}$. Therefore, this student should be matched to $s_{(1)}$ in all envyfree allocations, else $(i_1, s_{(1)})$ will form a blocking pair.

Now consider student $i_2$. If $s_{(2)}$ is the top choice of student $i_2$, then they should be matched with each other in any envyfree allocation, otherwise they would form a blocking pair since by definition, school $s_{(2)}$ has remaining capacity and student $i_2$'s priority qualifies them from one of the remaining seats. 
If student $i_2$ actually strictly prefers school $s_{(1)}$, the construction of the allocation through the application of the SMBP condition means that school $s_{(1)}$ has already reached its capacity with a higher priority student and is therefore not a feasible assignment for student $i_2$ as part of an envyfree allocation. Given this, assigning student $i_2$ to any other school than $s_{(2)}$ would result in a blocking pair because, by definition, student $i_2$ prefers $s_{(2)}$ to all other schools $s_{(k)}$, $k\geq2$, and student $i_2$'s priority qualifies them for one of the remaining seats.    

The same reasoning applies to student $i_3$ and the subsequent students. Each student $i_k$ will be matched to their respective school $s_{(k)}$ in every envyfree allocation. If $s_{(k)}$ is not their top choice, their preferred schools have already filled with higher priority students. This establishes that there exists a unique envyfree allocation in the market. 

Next, we show that the unique stable matching is efficient. Note that student $i_1$ is assigned to their top choice. If student $i_2$ is not assigned to their top choice, the only way to improve them is to change the assignment to $s_{(1)}$, but then $i_1$ will be worse off. In the same way, if $i_3$ is not assigned to their top choice, the only way to improve their situation is to assign them to $s_{(1)}$ or $s_{(2)}$, but then $i_1$ or $i_2$ will be worse off. Repeating the argument for the rest of the students establishes that this unique envyfree allocation is also efficient. 

The claim follows from the fact that DA, the school-proposing DA and IA all produce an envyfree allocation in equilibrium \citep{roth1984misrepresentation, ergin2006games}.
\end{proof}

Proposition \ref{GMBP} does not claim that DA produces the same allocation as TTC. The reason is that the simplification process can remove trading opportunities for TTC. However, if the original market satisfies the SMBP condition, then we can prove the stronger claim that TTC, student-proposing DA, school-proposing DA and IA, all produce the same allocation. In order words, the choice of the algorithm is second order.

\begin{proposition}[\textbf{Irrelevance of the algorithm}]\label{irrelevance}
If $\mathcal{E}=(\succ,\ P,\ q)$ satisfies the sequential mutually best pairs condition, then TTC, DA, the school-proposing DA and IA yield the same allocation and this allocation is both efficient and envyfree. 
\end{proposition}

\begin{proof} 
Consider the allocation produced by the reordering of students and schools associated with the definition of SMBP: $(i_1, s_{(1)})$, $(i_2, s_{(2)})$, ..., $(i_n, s_{(n)})$. Since $\mathcal{E}$ satisfies SMBP, this allocation corresponds to the DA outcome. Proposition 1 from \cite{dur2022characterization} shows that, for any school choice market, every student $i$ weakly prefers their assignment under TTC to any school $s$ for which $i$ is among the highest $q_s$ priority students. 
The proof proceeds by iteration. Take student $i_1$. Given that $i_1$ is among the top $q_{s_{(1)}}$ students in school $s_{(1)}$'s priority list, then $i_1$'s assignment under TTC is weakly preferred to $s_{(1)}$. As $s_{(1)}$ is $i_1$'s top choice, $i_1$ is assigned to $s_{(1)}$ under TTC. 

Now consider $i_2$. By definition, they are among the top $q_{s_{(2)}}-1_{\{s_{(2)} = s_{(1)}\}}$ students in school $s_{(2)}$'s priority list, so their assignment under TTC must be weakly preferred to $s_{(2)}$. By definition also, we know that $s_{(2)} \succeq s_{(k)}$, $k>2$. So by the application of Proposition 1 from \cite{dur2022characterization}, student $i_2$ must be assigned to school $s_{(2)}$ or (if $s_{(1)} \succ_{i_{2}} s_{(2)}$) to $s_{(1)}$ under TTC. We can rule out this second possibility since, by the previous argument, $i_1$ \textit{must} be assigned to $s_{(1)}$ under TTC and, if $s_{(2)} \neq s_{(1)}$, it means that school $s_{(1)}$ had only one seat.  
Continuing the process for each student, we can see that every student $i_k$ is assigned to $s_{(k)}$ under TTC, like under DA.
\end{proof}

A natural question that arises is to what extent GMBP is also necessary for DA to be efficient. The answer is negative. There are school choice markets where efficiency and envyfreeness are compatible, yet they do not satisfy GMBP as Example \ref{example:not-necessary} illustrates.

\begin{example}\label{example:not-necessary}
  Consider the following market with preferences and priorities as follows (all schools have unit capacity):
  \[
  \begin{array}{cccc}
  i_1: & s_3 & s_1 \\
  i_2: & s_2   \\
  i_3: & s_1 & s_2 & s_3 \\    
  \end{array}
  \hspace{50pt}
  \begin{array}{ccc}
  s_1: & i_1 & i_3 \\
  s_2: & i_2 & i_3 \\
  s_3: & i_3 & i_1 \\
  \end{array}
  \]
  \smallskip  
  
Note first that the school choice environment cannot be further simplified: there are no irrelevant schools in students' preferences. Furthermore, this market does not meet the generalized mutually best pair condition: while $i_2$ and $s_2$ are mutually best, the process stops there. 
\noindent Yet, DA is efficient and produces the same outcome as TTC, namely, $(i_1, s_3),  (i_2, s_2),  (i_3, s_1)$. 
\noindent Interestingly, this example is also one where the set of envyfree allocations is not unique: the school-proposing DA produces $(i_1, s_1),  (i_2, s_2),  (i_3, s_3)$. $\square$
\end{example}

It turns out that the non uniqueness of the envyfree allocation in Example \ref{example:not-necessary} captures exactly the characteristics of the school choice environments in which DA is efficient even though they do not satisfy GMBP. This leads us to derive the following result: 

\begin{proposition}\label{multiple}
School choice market $\mathcal{E}=(\succ,\ P ,\ q)$ satisfies the GMBP condition if and only if it admits a unique envyfree allocation and that envyfree allocation is efficient.
\end{proposition}

\begin{proof}
Given Proposition \ref{GMBP}, we only need to establish the ``if" part. So suppose $\mathcal{E}=(\succ,P\ ,q)$ admits a unique envyfree allocation, $\mu$, that also happens to be efficient. The proof proceeds by contradiction. Suppose that $\mathcal{E}$ does not satisfy the GMBP condition, i.e. the process of identifying mutually best pairs sequentially by applying Definition \ref{def:SMBP} to the simplified market $\mathcal{E^*}$ either does not start or ends before ordering all students.

Consider the remaining students and schools with their residual capacities after removing all existing mutually best pairs, if any. We will focus on this submarket for the remainder. With a slight abuse of notation, we let $\succ^*$, $P^*$ and $q^*$ represent the preferences, priorities and school capacities in that submarket.

In the simplified market (and thus in the submarket), students can only have one safe school, i.e. a school $s$ in which they appear among the top $q_s^*$ students, on their rank order lists, for otherwise one of these schools would be irrelevant. This means that the sets of priority students of different schools are disjoint and implies that the school-proposing DA finishes after the first round with every student receiving one offer at most and accepting it. Because, by assumption, there is a unique envyfree allocation, the resulting allocation corresponds to $\mu$.

Since $\mu$ is efficient, one of the students in the submarket must be assigned to their preferred school among those with remaining capacity. Call this student $i_1$ and the school to which they are assigned under $\mu$, $s_1$. Then, $i_1$ and $s_1$ are mutually best pairs, a contradiction with the assumption that there were no more mutually best pairs in the submarket. 
\end{proof}

An implication of Proposition \ref{multiple} is that GMBP is the most general condition to capture the type of preference-priority congruence that will ensure that DA is efficient. The other cases when DA is efficient are cases where the set of envyfree allocations is not a singleton and the specific structure of preferences and priorities are such that competition for schools is low enough that DA does not lead to (too much) efficiency-destroying rejection chains. These will also be cases for which the choice of the algorithm matters.

\section{The GBMP condition in stylized school choice environments}\label{relevance}
This section discusses, in the context of stylized school choice environments, how our condition compares with existing conditions that guarantee the efficiency of DA. The exercise helps crystallize the relationship among existing conditions and build intuition for the way in which our condition enlarges the set of known environments for which DA is efficient. In the next section, we complement these results with simulations. 

To do this, we turn to cardinal representations of preferences and priorities. Let $u_{is}$ denote a cardinal utility representation for student $i$’s preference for school $s$, with the convention that $s \succ_{i} s'$ if and only if  $u_{is}> u_{is'}$. Likewise, let $\pi_{is}$ describe student $i$'s priority at school $s$ with the convention that $i P_s j$ if and only if $\pi_{is}>\pi_{js}$. 

Table \ref{tab:compare} shows to what extent different school choice environments are covered by existing conditions and our GMBP condition. We start with \cite{ergin2002efficient} who was the first to investigate when DA is efficient in a school choice context. He identified acyclicity of priorities as a necessary and sufficient condition for DA to be efficient for any profile of preferences.\footnote{Acyclicity requires (in addition to a scarcity condition) that there are no three students $i,j,k$ and two schools, $s$ and $s'$, such that
$i P_s j P_s k P_{s'} i$. \cite{ergin2002efficient} shows that this is equivalent to requiring, that for every pair of schools and for every student ranked below the sum of the two schools’ capacities in one school, that student’s position differs at most by one across the two schools' lists of priorities. \cite{ehlers2010efficient} and \cite{erdil2019efficiency} extend this result to the case where priorities are coarse and to quotas, respectively.} While his condition does not place any restriction on preferences,  it is restrictive. Of all the stylized environments we consider in Table \ref{tab:compare}, it is only met in school choice environments where all schools use the same priorities, e.g., because seats are allocated on the basis of a test score or on the basis of a single tie-breaking rule.\footnote{Such priorities are common in university admissions, e.g. in China \citep{chen2017chinese}, Germany \citep{grenet2022preference}, Spain \citep{arenas2022gender}  and Turkey \citep{arslan2022empirical}, but also in secondary education, e.g. in Mexico \citep{chen2019self}, Romania \citep{pop2013going} and Singapore \citep{teo2001gale} or for exam schools in many countries.} 

All other conditions identified in the literature happen to rely on the existence of a mutually best pair condition, either in their definitions or in their proofs.\footnote{This includes \cite{alcalde1994exchange}, \cite{clark2006uniqueness}, \cite{eeckhout2000uniqueness}, \cite{legros2010co}, \cite{niederle2009decentralized}, \cite{reny2021simple}, \cite{rong2020stable}, and \cite{salonen2018mutually}. } Where the conditions differ is in the domain over which this mutually best pairs condition is required to hold. 

The strongest condition requires the mutually best pair condition to hold on the market and on every submarket where some students are removed and priorities are adjusted accordingly, or some schools are removed and rank order lists are adjusted accordingly, or both. In Table \ref{tab:compare}, we refer to this condition as ``MBP everywhere". For one-to-one matching environments, it comes down to \cite{alcalde1994exchange}'s $\alpha$-reducibility condition as adapted by \cite{clark2006uniqueness} and, equivalently, to \cite{niederle2009decentralized}’s aligned preferences.  In many-to-one environments, it is equivalent to \cite{salonen2018mutually}’s single-peaked preferences.

Environments where school priorities and student preferences depend on the same student-school match-specific component (e.g. distance, religion, academic inclination) satisfy this condition. Such preferences and priorities take the form $u_{is} = \pi_{is} = d_{is}$, where $d_{is}$ represents the student-school ``match quality". The condition also holds if, in addition, schools value student quality or students value school quality as long as they value these characteristics identically (no heterogeneity). In this case, preferences take the form $u_{is} = d_{is} +v_s$, and priorities take the form $\pi_{is} = d_{is}+g_i$. 
To verify that this environment satisfies the MBP condition everywhere, consider a $I \times N$ matrix, with elements $\phi_{is}=d_{is}+v_s+g_i$. The matrix rows represent students’ preferences. Indeed, $\phi_{is} > \phi_{is'}$ if and only if $u_{is}=d_{is}+v_s>u_{is'}=d_{is'}+v_{s'}$ (the $g_i$ term drops out). Likewise, the matrix columns represent school priorities: 
$\phi_{is}>\phi_{i's}$ if and only if $\pi_{is}=d_{is}+g_i>\pi_{is}=d_{i's}+g_{i'}$ (the $v_s$ term drops out). Assuming no identical student-quality match quality, a property of this matrix is that there is always an element that is maximal, for every submatrix. In other words, the MBP condition holds everywhere.  

Note that the richer and more realistic environments where students are heterogeneous in their valuation of quality (e.g. \citealp{abdulkadirouglu2017welfare}) do not meet any of the conditions generically. For example, if $u_{is} = d_{is} +\alpha_i v_s$, where $d_{is}$ stands for the inverse distance  and priorities are distance-based $\pi_{is} = d_{is}$, we can easily construct a situation where a student prefers the school further away because they value quality more, thereby blocking the construction of a mutually best pair. Specifically, take two students, $i$ and $j$, and two schools, $s$ and $s'$. Assume that student $i$ values school quality more than student $j$, i.e. $\alpha_i>\alpha_j$. Further assume that school $s$ has a higher quality than school $s'$, $v_s>v_{s'}$. Finally, assume that $d_{is'} > d_{js'} > d_{js} > d_{is}$ (if students live on a street that goes from one school to the other, it means that they both live closer to school $s'$ than to school $s$, but student $i$ even more so). If student $i$'s preference for quality is sufficiently strong to prefer school $s$, then $u_{is} = d_{is}+\alpha_i v_s > u_{is'}=d_{is'}+\alpha_i v_{s'}$ while $u_{js'}>u_{js}$ and $iP_{s'}j$ and $jP_si$. There is no mutually best pair in this submarket and therefore this environment does not satisfy the MBP everywhere condition.

Another environment where the MBP condition holds everywhere is when students prefer the school where their sibling, if any, goes, and where schools prioritize students with siblings and use a single tie-breaking rule for all others (assume for simplicity that the number of siblings benefiting from the priority is less than the number of seats available). To see this, consider any subset of schools and students. If there is a student with a sibling in one the schools, then they will be mutually best pairs. If no student has a sibling, then all schools will be ranking students in exactly the same way, and one of these schools (at least) will be the preferred choice of one of the students (we are back in the environment of the first row of Table \ref{tab:compare}).  

\begin{table}[htp]
\centering
    \caption{Comparing conditions across school choice environments}
    \label{tab:compare}
\scalebox{0.75}{

\begin{tabular}{L{2.9cm}L{2.9cm}L{3cm}cL{2.9cm}L{2.8cm}L{2cm}}
\toprule
Preferences                                                   & Priorities & Application & Ergin acyclity & MBP everywhere & SMBP & GMBP  \\
\midrule
Any          & Identical priorities (e.g. based on test scores)          &       University and high school admissions in several countries       &  $\checkmark$              &        $\checkmark$        &     $\checkmark$           &  $\checkmark$      \\
\midrule
$u_{is} = d_{is}$      &   $\pi_{is}=d_{is}$         &         Match-quality priorities and preferences    &                &          $\checkmark$       &           $\checkmark$      &    $\checkmark$    \\
\midrule
$u_{is}=d_{is}+v_s$  &    $\pi_{is}=d_{is} + g_i$          &       Match-quality + common priorities and preferences      &                &          $\checkmark$            &   $\checkmark$                   &    $\checkmark$         \\
\midrule
Prefers school with sibling, no restriction otherwise &         $\pi_{is} = \boldsymbol{1}_{\{i = \text{sibling}\}} + \varepsilon_{i} $ ($\varepsilon_{i} \in [0,1]$)  &    Sibling priorities, single tie-breaking rule for rest     &                &           $\checkmark$     &     $\checkmark$           &    $\checkmark$   \\
\midrule
Prefers one of their catchment schools, with some exceptions in run-away areas            &     $\pi_{is} = \boldsymbol{1}_{\{i = \text{in catchment}\}} + \varepsilon_{i} $ ($\varepsilon_{i} \in [0,1]$)        &    Guaranteed admission in catchment area, single tie-breaking otherwise    &        &   $\checkmark$             &   $\checkmark$         &    $\checkmark$   \\
\midrule
Prefers one of their catchment schools, with some exceptions          &     $\pi_{is} = \boldsymbol{1}_{\{i = \text{in catchment}\}} + \varepsilon_{i} $ ($\varepsilon_{i} \in [0,1]$)        &    Guaranteed admission in catchment area, single tie-breaking otherwise    &        &              &   possible         &    more likely   \\

\bottomrule

\end{tabular}
}
\smallskip

\begin{minipage}{1\textwidth}
       \scriptsize 
        \textbf{Notes:} The MBP condition everywhere is satisfied by \cite{alcalde1994exchange}'s $\alpha$-reducibility condition, \cite{niederle2009decentralized}’s aligned preferences, and (in multi-unit capacity environments) to \cite{salonen2018mutually}’s single-peaked preferences. The SMBP condition is satisfied by the environments for which \cite{salonen2018mutually}'s Iterated Best Match process converges. 
    \end{minipage}
\end{table}

The second strongest condition requires the mutually best pair condition to hold sequentially. In one-to-one settings, this is \cite{eeckhout2000uniqueness}’s condition. In one-to-many settings, it corresponds to environments for which \cite{salonen2018mutually}’s Iterated Best Match process converges to a non-wasteful matching. The advantage of such condition relative to the MBP condition everywhere is that it ignores unrealistic preferences that some students may have (schools on the top of their rank order lists that they have no chance of getting) and that prevent the MBP condition to hold. Indeed, as part of the sequential process of matching mutually best pairs, these schools are likely to fill up and we can then focus on their more realistic choices when searching for mutually best pairs.

The fifth and sixth rows of Table \ref{tab:compare} provide an illustration of the type of environments this condition allows for. In the example, the area is divided into catchment areas, with multiple schools each. Priority is given to students living in the catchment area and these students have essentially guaranteed access to at least one of them (in case of excess demand for one school, a single tie-breaking rule is used). Assume that most students prefer to go to a school in their catchment area but some prefer a school outside of their catchment area. 

If the catchment areas that attract out-of-catchment first choices are distinct from the catchment areas from which the students with these preferences come from, e.g. because there are popular catchment areas and run-away catchment areas, then MBP is still satisfied everywhere (fifth row). To see this, take any two students and the schools which they list first. If at least one of them makes a within-catchment-area choice, then at least one of them has top priority at their preferred school. If both make an out-of-catchment-area choice, the same priority ordering applies to them because of the single tie-breaking rule and therefore at least one of them is part of a mutually best pair. 

As soon as there is one student in one catchment area (say $A$) who prefers a school in another catchment area (say $B$), and the reverse, the ``MBP condition everywhere" fails (just take the submarket made of these two students and the two schools they prefer). 

However, it may still satisfy the SMBP. Indeed, through the sequential match of mutually best pairs, some schools, which the students making out-of-catchment-area choices listed, will reach capacity and will therefore exit the consideration set for these students. The sequential process will converge to a reduced set of students making out-of-catchment-area choices. This school choice market will satisfy the SMBP if there are no cross-catchment-area choices. To take an extreme example, simply suppose that these schools which attract out-of-catchment-area choices are so popular that they fill up with students from their own catchment area. This market will satisfy the SMBP condition.       

We can use this example to illustrate what the GMBP condition allows for, on top of the SMBP condition. Consider a student (say student $a$) in catchment area $A$ who lists a school in catchment area $B$ as first choice. That student will be ranked below every student in catchment area $B$ who listed that school as acceptable, even if they are not interested in that school because they are sure to get another school in catchment area $B$ if they ask for it (safe school). These students are hampering the formation of mutually best pairs and the GMBP, by truncating students' preferences at their safe school and updating priority lists accordingly, will free up space on the priority list of schools in area $B$ and make the formation of mutually best pairs easier. This benefit comes on top of what SMBP already delivered, namely the elimination of unrealistic choices from students' preferences, following the removal of mutually best pairs. This explains the ``more likely" entry in Table 2. 

We are unaware of generic environments where the GMBP condition is always met but the SMBP condition is not. However, this example illustrates how GMBP enables the identification of an expanded set of school choice environments where DA is efficient. The next section quantifies this statement by evaluating, in a parametric family of school choice environments, the fraction of markets (realizations of preferences and priorities) for which DA is efficient, the SMBP condition is satisfied, and the GMBP is satisfied. 

\smallskip

\section{Quantification}\label{sec:simulation}

Having shown qualitatively how the GMBP condition expands on existing conditions, we turn in this section to a quantitative assessment of the GMBP condition. Specifically, we generate a large number of school choice markets and check (1) to what extent DA is efficient in those markets, and (2) to what extent these markets satisfy the sequential and generalized mutually best pair conditions. 

Building on the recent empirical literature in school choice that estimates preferences for schools (e.g. \citealp{hastings2009heterogeneous, abdulkadirouglu2017welfare,calsamiglia2020structural,fack2019beyond, pathak2021well}), we assume that cardinal utilities underlying student preferences take the following form:
\[
u_{is} = \lambda \left(\delta d_{is}+ (1-\delta)v_s \right) + (1-\lambda)\varepsilon_{is},
\]
where, as before, $d_{is}$ stands for the student-school match quality (distance, religion, academic inclination), $v_s$ captures characteristics of the school that are valued equally by all students, and $\varepsilon_{is}$ is an idiosyncratic component capturing individual taste.\footnote{The main difference with empirically estimated preferences is that our coefficients on the match quality and school characteristics are assumed to be common across students.} 

The first term, $\delta d_{is}+ (1-\delta)v_s $, captures the structural part of preferences, driven by match quality and school characteristics. When $\delta = 1$, preferences are driven by match quality, which can be understood as the result of horizontal differentiation between schools, whereas when $\delta = 0$, schools are vertically differentiated in the eyes of students. The parameter $\lambda$ determines to what extent idiosyncratic preferences matter. When $\lambda = 1$, there are no idiosyncratic factors beyond match quality. When $\lambda = 0$, preferences are entirely idiosyncratic and independent of priorities. 

We consider the following priority structure for schools: 
\[
\pi_{is} = \alpha \left(\beta d_{is} + (1-\beta) g_i \right) + (1-\alpha)\eta_{is},
\]
where $g_i$ captures priorities based on student characteristics and single tie-breaking, and $\eta_{is}$ captures residual priorities based on idiosyncratic factors or multiple tie-breaking. The parameter $\alpha$ measures the degree of structure on priorities, whereas the parameter $\beta$ measures to what extent priorities are driven by match quality rather than student characteristics valued equally by all schools (e.g. grades). 

We carry out simulations as follows. For each school market environment -- characterized by the vector of parameters $(\lambda, \delta, \alpha, \beta)$, the number of students $n$, the number of schools $m$, and the school capacities $q$ (we assume all schools have identical capacities), we draw 1,000 realizations of the vector of variables $(d_{is}, v_s, \varepsilon_{is}, g_i, \eta_{is})$, where all variables are independently drawn from the uniform distribution on $[0,1]$. This generates 1,000 market realizations for every school market environment. We set $n = 1,000$, $m = 50$ and $q = 20$.\footnote{We played with different market sizes and school sizes, with no qualitative changes in the results.} 

\begin{table}[htp]
    \centering
      \caption{Percentage of markets where DA is efficient, SMBP and GMBP are satisfied}
    \label{tab:summary}
    \footnotesize
       \begin{tabular}{llccc}
    \toprule
     &   &   (1)   &  (2)   &   (3) \\
     \cmidrule{3-5} 
     Preferences & Priorities &   DA is efficient& SMBP & GMBP\\
     \midrule
      \multirow{3}{*}{$\lambda=1$} & $\alpha=1$ &100  &  100 &  100  \\
                            & $\alpha=0.95$ &  73.86&	64.43&	72.25   \\
                            & $\alpha = 0.90$ &    15.55	&12.24	&15.05         \\ 
                          \midrule 
   \multirow{3}{*}{$\lambda=0.75$} & $\alpha=1$ & 15.94 &	8.70 &	14.78   \\ 
                                    & $\alpha=0.95$ &  4.67 & 0.01 & 3.42\\ 
                                    & $\alpha=0.90$ & 0.23 &	0.00 	& 0.13\\    
                                    \midrule
    \multirow{3}{*}{$\lambda=0.5$} & $\alpha=1$ &  8.08 &	4.82 	& 6.57  \\
                                    & $\alpha=0.95$ & 1.49 & 0.00 	& 0.41  \\ 
                                    & $\alpha=0.90$ & 0.08 &	0.00 &	0.01 \\
                                    \midrule 
      \multirow{3}{*}{$\lambda=0.25$} & $\alpha=1$ & 6.23 &	4.76 	& 5.08  \\
                                    & $\alpha=0.95$ & 1.02	&0.00	&0.10    \\ 
                                    & $\alpha=0.90$ & 0.06	&0.00	& 0.00 \\ 
                                    \midrule 
      \multirow{3}{*}{$\lambda=0$} & $\alpha=1$ &  5.17 	&4.76 & 	4.79   \\
                                    & $\alpha=0.95$ & 0.62 &	0.00 	& 0.07 \\ 
                                    & $\alpha=0.90$ & 0.07	&0.00	& 0.00\\ 
      \bottomrule
    \end{tabular}
  \begin{minipage}{0.6\textwidth}
  \medskip 
  
  \footnotesize
  \textbf{Notes:} The numbers indicate the percentage of school choice markets for which DA is efficient (column (1)) and that satisfy, respectively, the SMBP condition (column (2)) and the GMBP condition (column (3)). Percentages are computed first on the basis of 1,000 independent draws of variables $(d_{is}, v_s, \varepsilon_{is}, g_i, \eta_{is})$ from the uniform distribution $[0, 1]$, for a given value of $\delta$ and $\beta$, then averaged over the  $\delta$ and $\beta$ parameters taking values from 0 to 1 in 0.05 increments. 
  \end{minipage}
\end{table}

Table \ref{tab:summary} reports the percentage of market realizations where DA was found to be efficient and which satisfied the SMBP and GMBP conditions, respectively. We fix the values for $\lambda$ and $\alpha$, and average over all market realizations and values for $\delta$ and $\beta$, where $\delta$ and $\beta$ vary from 0 to 1 in 0.05 increments. The table presents the results for $\lambda$ taking the values of $1,\ 0.75,\ 0.5,\ 0.25$, $0$, and $\alpha$ taking the values of $1$, $0.95$,  $0.9$.\footnote{We only present results for higher values of $\alpha$ because the fraction of environments where DA is efficient sharply drops with $\alpha$.}  

Three observations stand out from the table. First, the ability of DA to generate efficient allocations varies strongly across environments. Except when $\lambda=1$ and $\alpha=1$, which corresponds to the third row of Table \ref{tab:compare}, DA is not efficient for some realizations for preferences and priorities. The ability of DA to be efficient drops sharply with students' idiosyncratic preferences (low $\lambda$) and schools' idiosyncratic priorities (low $\alpha$). Idiosyncratic preferences (on both sides) is when the trade-off between respecting preferences and respecting priorities is largest. 

\begin{figure}[hp]
\centering
   \caption{Share of markets where DA is efficient (green), the SMBP is satisfied (blue), and GMBP is satisfied (red), as a function of $\delta$ and $\beta$}\label{fig:sim}
     \begin{subfigure}[b]{0.53\textwidth}
         \includegraphics[scale=0.23]{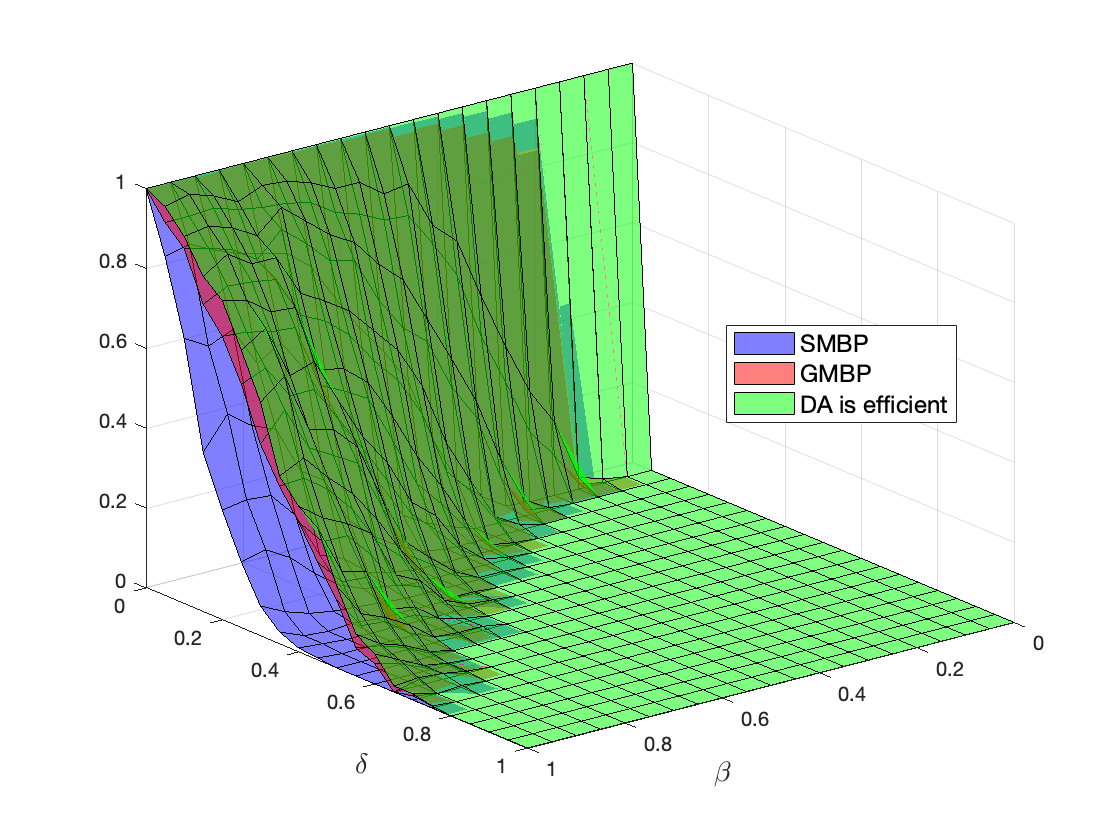}
         \caption{$\alpha=1,\lambda=0.75$}
         \label{fig:2}
     \end{subfigure}
     \begin{subfigure}[b]{0.45\textwidth}
         \includegraphics[scale=0.25]{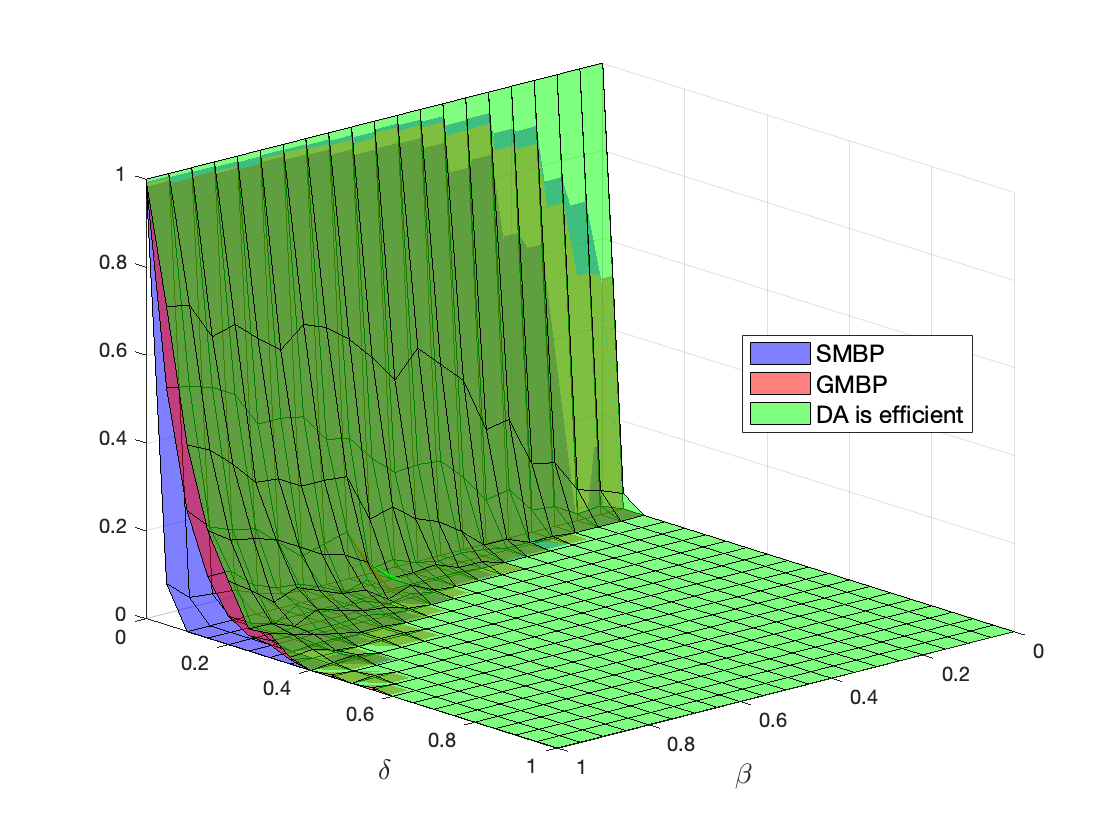}
         \caption{$\alpha=1,\lambda=0.5$}
         \label{fig:3}
     \end{subfigure}
       \hfill
         \begin{subfigure}[b]{0.5\textwidth}
         \includegraphics[scale=0.23]{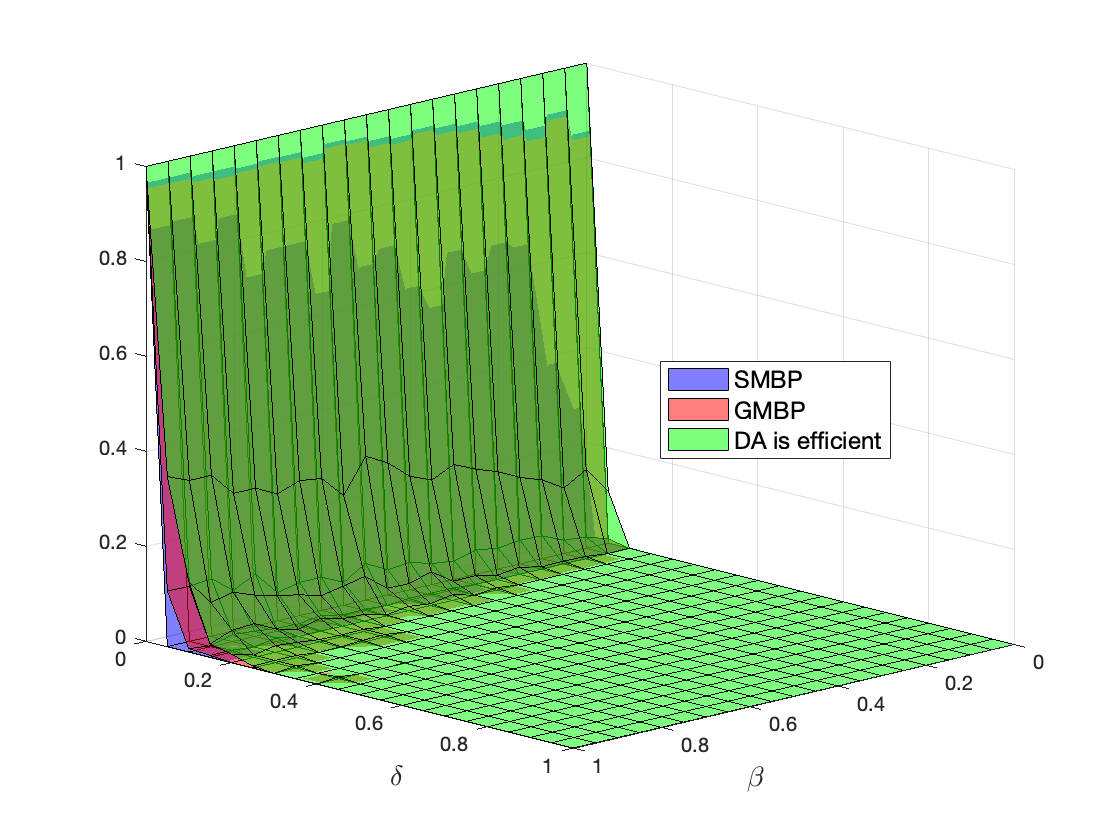}
         \caption{$\alpha=1,\lambda=0.25$}
         \label{fig:4}
        \end{subfigure}
        \hfill
        \begin{subfigure}[b]{0.45\textwidth}
         \includegraphics[scale=0.25]{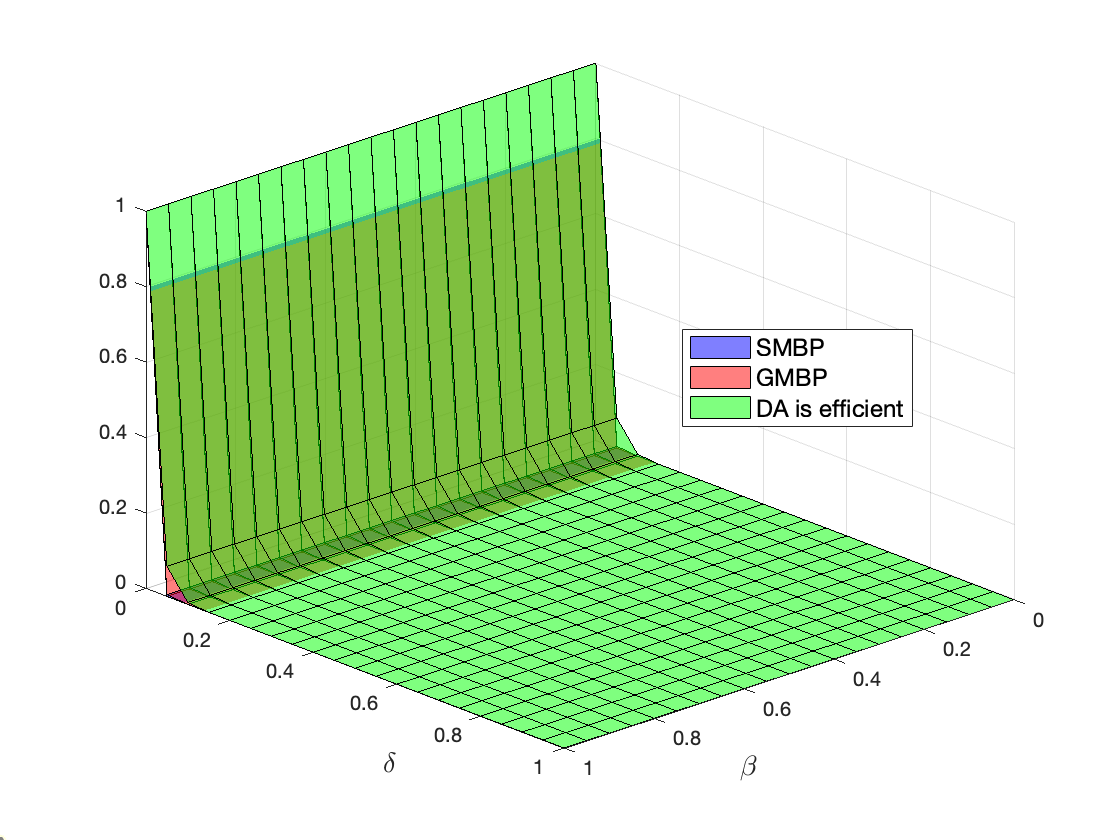}
         \caption{$\alpha=1,\lambda=0$}
         \end{subfigure}
\begin{minipage}{1\textwidth}
        \footnotesize
        \textbf{Notes:} Simulations for a school market with 1,000 students, 50 schools and school capacities of 20. Shares are averages over 1,000 independent draws of variables $(d_{is}, v_s, \epsilon_{is}, g_i, \eta_{is})$, for a given value of the $\lambda$, $\alpha$, $\beta$, and $\delta$ parameters. $\delta$ and $\beta$ vary from 0 to 1 in 0.05 increments. $\alpha$ is set to 1 (no idiosyncratic component on priorities), and $\lambda$ takes the values of 0.75,\ 0.5,\ 0.25,\ 0.
        \end{minipage}
\end{figure}

Second, the GMBP condition systematically captures a larger fraction of the environments for which DA is efficient than the SMBP condition. The gap is especially large as soon as we allow for idiosyncratic preferences ($\lambda < 1$) or priorities ($\alpha<1$). Recall from Example \ref{example:truncation} that the GMBP condition is able to identify environments where DA is efficient, which the SMBP did not, when removing the irrelevant schools for students leads to new mutually best pairs. This situation is more likely to occur when school priorities and student preferences go in opposite directions: 
schools give higher priority to students who prefer them the least. Smaller values for $\lambda$ or $\alpha$ allow for this possibility.

The third insight is that, for a given value of $\lambda$, the GMBP condition is able to identify a large proportion of the school choice environments for which DA is efficient, especially when $\alpha$ takes high values. From Proposition \ref{multiple}, we know that the remaining cases are cases where there are multiple envyfree allocations.

The numbers in Table \ref{tab:summary} are averaged over the different values of $\delta$ and $\beta$. Figure \ref{fig:sim} provides a closer look at how these numbers vary over $\delta$ and $\beta$. In all cases, we let $\alpha=1$, so priorities take the form $\pi_{is}=\beta d_{is}+(1-\beta)g_i$: schools care about match quality (the kind that is reciprocated in preferences) and student characteristics. The figures illustrate how the probability that DA is efficient quickly decreases with $\lambda$. 
Low values for $\delta$ (i.e. strong vertical differentiation of schools) and high values for $\beta$ (i.e. greater emphasis on match quality in school priorities) foster the efficiency of DA. A higher value of $\beta$ allows school priorities to ``align'' with preferences through the match quality, even when students value more factors, such as school quality. The GMBP condition is capable of identifying most of these markets when DA is efficient. 

\section{Discussion}\label{sec:conclusion}
When is there a trade-off between efficiency and envyfreeness or, equivalently, when is there a trade-off between preferences and priorities?  Our results confirm the conjecture put forward by \cite{pathak2017really} according to which ``correlation between preferences and priorities induced by proximity may, in turn, result in less scope for Pareto-improving trades across priority groups that involve situations of justified envy. This pattern may then result in a small degree of inefficiency in DA", and clarify to which extent the conjecture holds. Specifically, our generalized mutually best pairs condition maximally captures the set of environments where priorities and preferences are sufficiently congruent that DA is efficient, and there is no trade-off between efficiency and envyfreeness. Our condition is more general than \cite{pathak2017really}'s correlation conjecture between preferences and priorities because it restricts attention to those parts of preferences and priorities that are relevant for an envyfree allocation.

Our results shed light on the empirical evidence presented in Table \ref{tab:Evidence}. A small trade-off was found in (primarily) elementary school markets with priorities to siblings, staff, and some measure of distance (Boston, Ghent, New Orleans). The elementary school level is exactly the education level where one would also expect parents to place greater emphasis on proximity or selecting the same school as the older sibling in order to facilitate home logistics when children are not yet capable of travelling by themselves.\footnote{Most empirical studies of parental preferences restrict attention to one educational level and so there is little evidence for this comparative statement. \cite{harris2023schools} is an exception. Their data cover submitted preferences for both the elementary and high school levels in New Orleans. They find stronger preferences for sibling presence and distance for elementary school applications.} In other words, markets where preferences and priorities are congruent. Such markets are well captured by the functional forms in rows 2 and 3 of Table \ref{tab:compare} and high values of $\lambda$ and $\alpha$ in Table \ref{tab:summary}. 

On the other hand, the secondary and high school markets of Budapest and New York City, respectively, are characterized by a higher level of idiosyncratic school-specific priorities and, presumably, a higher level of horizontal differentiation across schools (different specialisation tracks). Our numerical results suggest that DA is less likely to be efficient in such markets. The high level of justified envy found in the data suggests that the trade-off between envyfreeness and efficiency might indeed be big in those school markets. 

Preference-priority congruence, as captured by the GMBP condition, does not capture the universe of environments where DA is efficient, however. There are situations, such as the one illustrated in Example \ref{example:not-necessary} where DA is efficient, yet the GMBP condition is not satisfied.  This happened because, while preferences and priorities conflicted, students had sufficiently different preferences that they could nevertheless get their first choice. Priorities were toothless. Our Proposition \ref{multiple} establishes that all these environments are characterized by a multiplicity of envyfree allocations where, therefore, the choice of the algorithm matters.


What lessons can policy-makers draw from our analysis? A first general lesson is to understand their markets -- what drives student preferences -- and assess to what extent school priorities are likely to be congruent with those, for the parts of the market where there is excess demand. If this is the case, the choice of the algorithm is likely to be second order. If not, extensive evaluation of different designs might be useful. A second lesson is that they can use their discretionary power, when available, to increase the probability that their markets meet the GMBP condition. An obvious example is the choice of the tie-breaking rule when priorities are weak. Our results here echo \cite{ashlagi2020matters}'s recommendation that popular schools use a single tie-breaking rule: fostering preferences and priority congruence is especially valuable for schools with excess demand. 



\newpage
\bibliographystyle{chicago}
 \bibliography{stability}

\newpage
\appendix
\section*{Appendix \label{sec:app}}

\subsection*{Description of the mechanisms}

\subsubsection*{Student-proposing Deferred Acceptance (DA)}

\noindent \textbf{Step 1:} Each student $i$ proposes to
the best school according to $\succ_{i}$. Each school $s$ provisionally accepts the $q_{s}$-highest ranked students,
according to $P_{s}$, among those students that have proposed to $s$, and
rejects the others.\smallskip

\noindent \textbf{Step $k$:} Each student $i$, who has
not been previously accepted, proposes to the best school according to $%
\succ_{i} $, among those schools that have not yet rejected $i$. Each school $s$ provisionally accepts the $q_{s}$-highest ranked students,
according to $P_{s}$, among those students that have proposed to $s$ along
steps 1 to $k$, and rejects the others.\smallskip

The algorithm terminates at the step where no rejections are made and
provisional acceptances become definitive by matching each school $s$ to the set of students provisionally accepted at this step.

\subsubsection*{School-proposing Deferred Acceptance (DA)}

\noindent \textbf{Step 1:} Each school $s$ proposes to
the $q_s$-highest ranked students according to $P_s$. Each student $i$ provisionally accepts the best school
according to $\succ_i$, among those schools that have proposed to $i$, and
rejects the others.\smallskip

\noindent \textbf{Step $k$:} Each school $s$, which has
been previously rejected, proposes to the next highest priority students according to $
P_s$ up to capacity $q_s$, among those students that have not yet rejected $s$. Each student $s$ provisionally accepts the best school,
according to $\succ_i$, among those schools that have proposed to $i$ along
steps 1 to $k+1$, and rejects the others.\smallskip

The algorithm terminates at the step where no rejections are made and
provisional acceptances become definitive by matching each school $s$ to the set of students provisionally accepted at this step

\subsubsection*{Top Trading Cycles (TTC)}

\noindent \textbf{Step 1:} Each student $i$ points to
the best school according to $\succ_{i}$. If no school is acceptable, $i$ points
to $s_{m+1}$ and 
is removed from the
problem.
Each school $s$ points to the best student according to $P_{s}$, and $s_{m+1}$ points to all students.

Since the sets of students and schools are finite, there exists at least one
cycle which is of the form $i_{1}\longrightarrow s_{1}\longrightarrow \cdots
\longrightarrow i_{K}\longrightarrow s_{K}\longrightarrow i_{1}$ or $%
i\longleftrightarrow s_{m+1}$, where $x\longrightarrow y$ means
\textquotedblleft $x$ points to $y$\textquotedblright.\footnote{
	There may be many cycles, although each student and school $s \neq s_{m+1}$ can be part of at most one cycle.}
Each student $i$ in a cycle is matched to the
school $s$ that they point to (if $i\longrightarrow s$),
in which case $i$
and a seat in $s$ are removed from the problem. If $i$ points to $s_{m+1}$. They remain 
unmatched and $i$ is removed from
the problem.
\smallskip

\noindent \textbf{Step $k$.} Each remaining student $i$ points to the best school according to $\succ_{i}$,
among the schools that still have empty seats.
Each school $s$ with an empty seat, points to the best student, according to
$P_{s}$, among the remaining students, and $s_{m+1}$ points to all of these students.
There is at least one cycle. Each student $i$ in a cycle of the form $%
i_{1}\longrightarrow s_{1}\longrightarrow \cdots \longrightarrow
i_{K}\longrightarrow s_{K}\longrightarrow i_{1}$ is matched to the school $s$
that they point to, and $i$ and a seat in $s$ are removed from
the problem. If $i$ points to $s_{m+1}$. One remains
unmatched and $i$ is removed from
the problem.
\smallskip

The algorithm terminates when each student $i$ is either matched to a school or to $s_{m+1}$. 
\bigskip

\subsubsection*{Immediate Acceptance (IA)}

\noindent \textbf{Step 1:} Each student $i$ proposes to
the best school according to $\succ_{i}$. Each school $s$ accepts the $q_{s}$-highest ranked students,
according to $P_{s}$, among those students that have proposed to $s$, and
rejects the others. Accepted students are definitive matched to the school. Schools' capacities are reduced by the number of students accepted. \smallskip

\noindent \textbf{Step $k$:} Each student $i$, who has
not been previously accepted, proposes to the best school according to $%
\succ_{i} $, among those available schools that have not yet rejected $i$. Each school $s$ accepts the $q_{s}$-highest ranked students,
according to $P_{s}$, among those students that have proposed to $s$, and rejects the others. Accepted students are definitive assigned to the school.Schools' capacities are reduced by the number of students accepted.\smallskip

The algorithm terminates at the step where no rejections are made. 
\bigskip

\end{document}